\documentclass[structabstract]{aa}
\usepackage{txfonts}
\usepackage{graphicx}
\usepackage{natbib}
\def\psrb{PSR~B1259-63}
\begin{document}

	\title{What caused the GeV flare of PSR B1259-63 ?}

	\author{G. Dubus\inst{1}
	\and 
	B. Cerutti\inst{2}}

	\institute{UJF-Grenoble 1 / CNRS-INSU, Institut de Plan\'etologie et d'Astrophysique de Grenoble (IPAG) UMR 5274, Grenoble, F-38041, France 
	\and
	Center for Integrated Plasma Studies, Physics Department, University of Colorado, UCB 390, Boulder, CO 80309-03960, USA}

    \date{Received ; Accepted }

\abstract{PSR B1259-63 is a gamma-ray binary system composed of a high spindown pulsar and a massive star. Non-thermal emission up to TeV energies is observed near periastron passage, attributed to emission from high energy $e^+e^-$ pairs accelerated at the shock with the circumstellar material from the companion star, resulting in a small-scale pulsar wind nebula. Weak gamma-ray emission was detected by the {\em Fermi}/LAT at the last periastron passage, unexpectedly followed 30 days later by a strong flare, limited to the GeV band, during which the luminosity nearly reached the spindown power of the pulsar. The origin of this GeV flare remains mysterious.}
{We investigate whether the flare could have been caused by pairs, located in the vicinity of the pulsar, up-scattering X-ray photons from the surrounding pulsar wind nebula rather than UV stellar photons, as usually assumed. Such a model is suggested by the geometry of the interaction region at the time of the flare.}
{We compute the gamma-ray lightcurve for this scenario, based on a simplified description of the interaction region, and compare it to the observations.}
{The GeV lightcurve peaks well after periastron with this geometry. The pairs are inferred to have a Lorentz factor $\approx 500$.  They also produce an MeV flare with a luminosity $\approx 10^{34}$\,erg\,s$^{-1}$ prior to periastron passage.  A significant drawback is the very high energy density of target photons required for efficient GeV emission.}
{We propose to associate the GeV-emitting pairs with the Maxwellian expected at shock locations corresponding to high pulsar latitudes, while the rest of the non-thermal emission arises from pairs accelerated in the equatorial region of the pulsar wind termination shock.}

\titlerunning{What caused the GeV flare of PSR B1259-63 ?}
\authorrunning{G. Dubus \& B. Cerutti}
   \maketitle

\section{Introduction}
\psrb\  is a 47.7 ms radio pulsar in a 3.5 year orbit around a Be star \citep{1992ApJ...387L..37J}. The pulsar has a high spindown power $\dot{E}\approx 8\times10^{35}$\,erg\,s$^{-1}$. An outburst of very high energy gamma ray (VHE, $>$100 GeV), X-ray, and radio emission occurs at periastron passage ($\tau\equiv0$\,d), when the pulsar wind interacts with the stellar wind and the equatorial disc surrounding the Be companion. A bow shock structure forms, whose appearance changes according to which component (stellar wind or Be disc) the pulsar wind interacts with. The $e^{-}e^{+}$ pairs present in the pulsar wind are isotropised and energised at the shock. The pairs radiate non-thermal emission as they flow away from the binary, producing a small-scale pulsar wind nebula (PWN).

High-energy gamma-ray emission (HE, $>$100 MeV) was detected for the first time during the last passage in 2010 \citep{2011ApJ...736L..10T,2011ApJ...736L..11A}. A weak $\approx 5\sigma$ ``brightening'' of the emission was detected starting at least 20 days before periastron ($\tau$-30\,d), ending a couple of weeks after ($\tau$+15\,d). The system was not detected in HE gamma rays for the next two weeks, and its  activity related to periastron passage was thought to be over when it surprisingly brightened in HE gamma rays at $\tau$+30\,d. This ``GeV flare'' reached a luminosity close to the pulsar spindown power, lasting nearly 2 months up to $\tau$+80\,d. The average spectrum was a power law of photon index $\Gamma$=1.4 with an exponential cutoff at 0.3 GeV, hardening with decreasing flux. The GeV flare was not accompanied by concurrent changes in emission in radio, X-rays, or VHE gamma rays \citep{2011ApJ...736L..11A,H.E.S.S.Collaboration:2013fk}, suggesting that the particles responsible for this GeV emission component have a narrow energy distribution, distinct from the power-law distributed pairs attributed to the shocked pulsar wind. 

The origin of the flare remains puzzling: how is the spindown power so efficiently converted to gamma rays and why does this occur suddenly one month after periastron passage ? The double-peaked shape of the radio and X-ray lightcurves has been associated with the pulsar ``crossings'' of the dense Be disc  \citep{Tavani:1997wv,Johnston:1999uq,2006MNRAS.367.1201C,2012ApJ...750...70T}, which is thought to be inclined with respect to the orbital plane \citep{1995MNRAS.275..381M}. However, the GeV flare starts $\approx$\,10 days after the post-periastron radio/X-ray peak, well after the presumed disc crossing times. The pulsar is far from the densest regions, being 60 to 100 $R_\star$ away from the star during the GeV flare (Fig.~1). Indeed, the radio pulsations, eclipsed by the circumstellar material starting at $\tau-16$\,d, turn back on two weeks prior to the GeV flare \citep{2011ApJ...736L..11A}. Hence, the mechanism for the GeV flare also appears distinct from that causing the radio/X-ray variability.

A high conversion efficiency to HE gamma-rays can be obtained if the pairs present in the pulsar wind (and its nebula) cool rapidly due to inverse Compton scattering. The companion star has a high luminosity so most models have considered the anisotropic upscattering of stellar photons \citep{Kirk:1999hr,2007MNRAS.380..320K,2009ApJ...702..100T,2011MNRAS.417..532P}.  The lightcurve peaks slightly before periastron, when the orbital geometry allows close to head-on scattering. This lightcurve is compatible with the brightening but not with the GeV flare. Taking the contribution from the Be disc does not change this conclusion since disc photons come predominantly from the region closest to the star, leading to the same lightcurve (\citealt{2011MNRAS.412.1721V,2012MNRAS.426.3135V}, Yamaguchi,  pers. comm.). \citet{2012ApJ...752L..17K} proposed that the flare is related to the rapidly changing shape of the bow shock as the pulsar exits the Be disc. The unshocked pulsar wind becomes unconfined along the direction of weakest external pressure, giving the cold pairs present in the wind more space (and time) to cool. A high energy density of seed photons with a favourable scattering geometry is still required: \citet{2012ApJ...752L..17K} postulated that this would be provided by local heating of the Be disc due to the pulsar crossing.

A possible clue is that the orbital phase of the GeV flare brackets inferior conjunction of the pulsar ($\tau$+60\,d, Fig.~1). The observational evidence presented above suggests the pulsar is away from the Be disc material, so that its ram pressure balances the stellar wind ram pressure. In this case, the bow shock is oriented towards the observer at inferior conjunction \citep{2011ApJ...736L..10T}. For parameters appropriate to \psrb, the wind momentum ratio $\eta$ is
\begin{equation}
\eta=\frac{\dot{E}/c}{\dot{M}_w v_w}\approx 0.5 \left(\frac{10^3\,\mathrm{km\,s}^{-1}}{v_w}\right) \left(\frac{10^{-8}\,\mathrm{M}_\odot\,\mathrm{yr}^{-1}}{\dot{M}_w}\right)\left(\frac{\dot{E}_w}{10^{36}\,\mathrm{erg\,s^{-1}}}\right).
\end{equation}
Semi-analytical approximations and numerical simulations show that the shock region asymptotes to a hollow cone far from the binary axis; its minimum ($\theta_{\rm in}$) and maximum ($\theta_{\rm out}$) opening angle  depend only on $\eta$ \citep[][and references therein]{2008MNRAS.387...63B,2011MNRAS.418.2618L}. For $\eta\approx0.5$, the shocked pulsar wind fills a cone from 50\degr\ to 65\degr, while the shocked stellar wind fills a cone from 65\degr\ to 110\degr\ \citep{2008MNRAS.387...63B}. With an orbital inclination $i\approx 30\degr$, the line-of-sight goes through the whole length of the shocked pulsar wind cone at inferior conjunction (Fig.~\ref{fcone}). Material in this cone has a speed $c/3$ immediately after the shock with the flow directed towards the observer, boosting the shocked pulsar wind emission due to relativistic effects \citep{2010A&A...516A..18D}. Hence, \citet{2011ApJ...736L..10T} and \citet{2012ApJ...753..127K} proposed that the GeV flare is due to Doppler-boosted synchrotron emission, providing an attractive explanation for its orbital phase. However, all co-located emission is impacted by Doppler boosting so the lack of simultaneous flaring at other frequencies (X-rays, VHE) is puzzling in this model. In addition, \citet{2012ApJ...752L..17K} commented that it would be a remarkable coincidence that the Doppler-boosted GeV luminosity happens to be nearly equal to the spindown luminosity.

Here, we investigate whether the GeV flare could have been due to inverse Compton scattering of photons from the shocked pulsar wind. The scattering geometry will be favourable if the emitting pairs are close to the pulsar and the seed photons originate from the ``cometary tail'' of shocked pulsar wind material.  The seed photons will then be back-scattered to the observer at inferior conjunction, when the shock cone sweeps the line-of-sight, providing a geometric explanation for the orbital phase of the GeV flare. Indeed, radio VLBI maps taken in 2010-11 show large changes in the position angle of the resolved emission from the shocked flow between periastron passage and $\tau+100$\,d (\citealt{Moldon:2012fk}, and in prep.). A detailed model would require a simulation of the interaction region coupled with radiative codes. For this exploratory work we computed the expected lightcurve using a simplified toy model, sufficient to discuss its consequences on the properties of the pairs in the pulsar wind and to point out a possible test: the detection of an MeV flare prior to periastron passage.
\begin{figure}
\centering
\resizebox{\hsize}{!}{\includegraphics{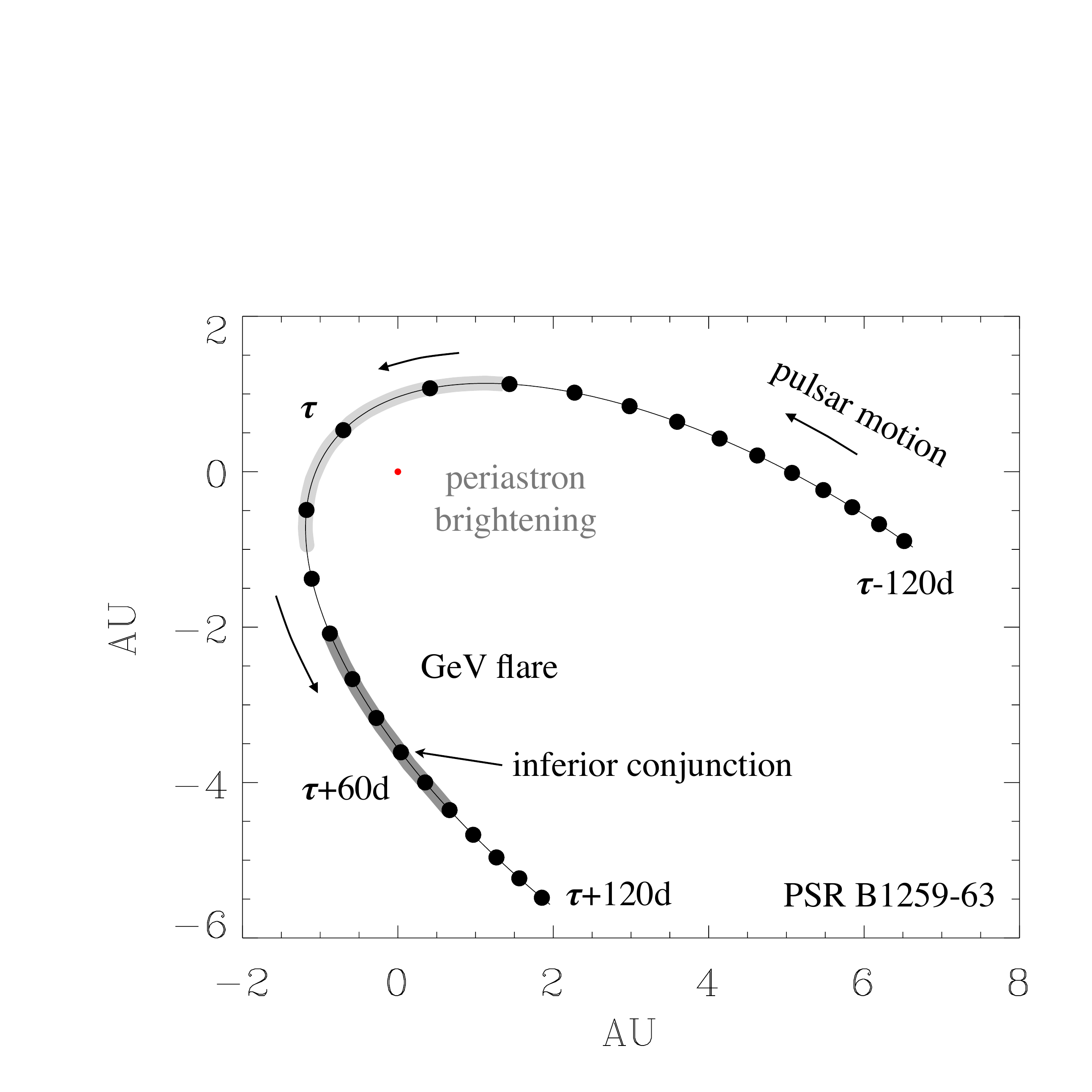}} 
\caption{Orbit of PSR B1259-63 close to periastron, projected assuming an inclination $i=30$\degr. Be star is the red dot at center (to scale). Black dots mark the pulsar position in 10 day intervals. The times of the GeV ``brightening'' and ``flare'' are highlighted.}
\end{figure}
\begin{figure}
\centering
\resizebox{\hsize}{!}{\includegraphics{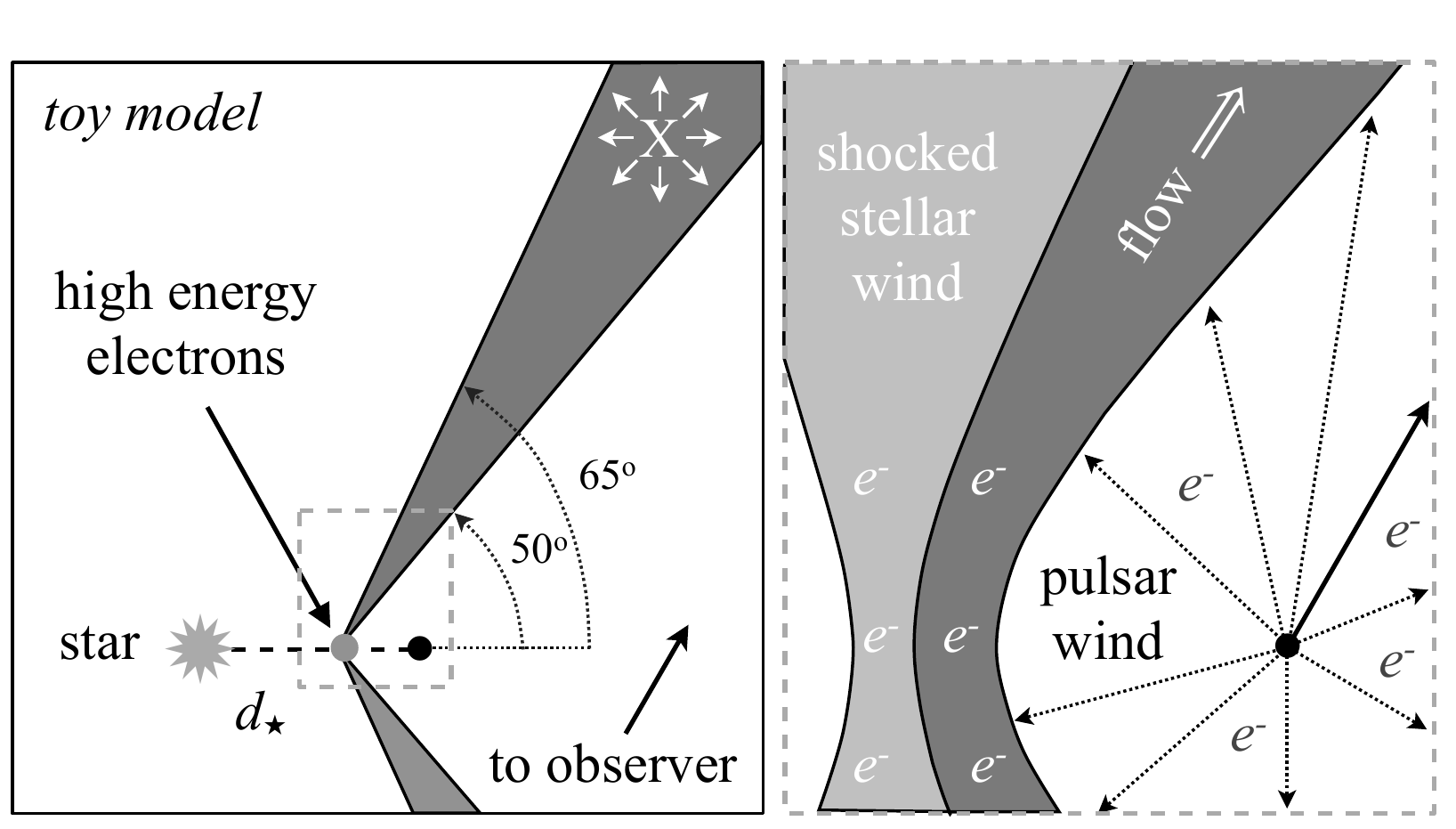}} 
\caption{Left: geometry of the toy model for inverse Compton scattering of X-ray emission from the cone of shocked pulsar wind (dark shaded region in both panels). Right: zoom on the region delimited by a dashed rectangle on the left, showing  possible sources of high energy electrons. The direction to the observer is $\approx60$\degr\ during the GeV flare (full line arrow in both panels).}
\label{fcone}
\end{figure}

\section{Model}
\subsection{Assumptions\label{model}}
We assume that the pulsar wind interacts only with the fast stellar wind at the time of the GeV flare (\S1) and approximate the shock region as a hollow cone, characterised by two opening angles measured from the cone axis $\theta_{\rm in}\approx 50$\degr\ and $\theta_{\rm out}\approx 65$\degr\ (see left panel of Fig.~\ref{fcone}). This hollow cone represents the shocked pulsar wind region and is the source of seed photons for inverse Compton scattering in our toy model. The particles in this cone are assumed to emit the seed photons isotropically (note, however, that the emission from the cone seen at a given location is not isotropic). The cone axis is oriented along the line-of-centres joining the Be star to the pulsar. The cone apex is located at the standoff distance between the two winds ($\approx 0.6\times$ the orbital separation from the Be star for $\eta$=0.5) although its precise location is not important here. 

This geometry does not provide a precise description of the interaction region close to the pulsar, where the shock width is finite and the respective location of the high-energy pairs and seed photons is complex {(see right panel of Fig.~\ref{fcone})}. The pairs responsible for the GeV emission could be localised in the pulsar wind, in the shocked pulsar wind, or in the shocked stellar wind. Here, we will assume that the pairs are located at the apex of the cone, and come back to their origin later (\S3.2).  Localising the pairs at the apex is reasonable if the seed photons are emitted on a much larger scale.

The pairs at the cone apex see a constant intensity $I_\nu$ of seed photons where the cone is filled in, and no intensity in the hollowed-out region or outside the cone. The radial distribution of emissivity in the cone is not needed to calculate the radiation energy density at the apex. As discussed in \S1, the GeV spectrum suggests a narrow energy distribution for the pairs. We took mono-energetic electrons of Lorentz factor $\gamma$.

The inverse Compton emission from the pairs is expected to be dominated by the upscattering of soft X-rays from the shocked pulsar wind. The inverse Compton power depends on the energy density of the seed photons: it  is maximum for interaction on X-ray photons since the observed spectral energy distribution breaks at a few keV \citep{2009ApJ...698..911U}. The Lorentz factor of the pairs must be $\gamma\approx 500$ to boost 1 keV photons to 300 MeV, where the spectrum cuts off. The inverse Compton interaction is at the limit between Thomson and Klein-Nishina regime, hence the cooling timescale is the smallest possible, maximizing the radiative efficiency. Other consequences of being close to the Klein-Nishina limit  are that (1) seed photons with energies $> 1$\,keV contribute little to the inverse Compton emission; (2) emission $\geq$ 1 GeV may be  absorbed due to pair production on the X-ray photons, depending on how the X-ray emission is distributed in the cone.

The soft X-ray emission is likely to arise in an extended region, larger than the zone where the pairs are located at the cone apex. Just like a PWN, the shocked pulsar wind is expected to radiate synchrotron emission from radio up to the radiative limit at $\approx 30$ MeV \citep[e.g.][]{1996ApJ...457..253D}, with the higher energy photons emitted from the regions closest to the apex of the termination shock \citep[e.g.][]{Kennel:1984gu,Dubus:2006lc}. Emission in X-rays involves particles that have significantly cooled and have been advected away from the shock apex. There is observational evidence that in LS 5039, where a comparable interaction is thought to occur, the X-rays are emitted far out of the system in an extended region \citep{Bosch-Ramon:2007fq,2011MNRAS.411..193S}. The shocked {\em stellar} wind may also contribute thermal X-rays: we have neglected it since the observations show that the non-thermal component dominates.

\subsection{Lightcurve}
The spectrum from the mono-energetic electrons is highly peaked at 300 MeV so the HE lightcurve follows closely the bolometric lightcurve. With the assumption described above, the bolometric lightcurve is straightforward to compute in the Thomson regime by following \citet{Henri:1997aa}. We find that the total power radiated by an electron of Lorentz factor $\gamma$ (velocity $\beta=v/c$) travelling at an angle $\mu=\cos \theta$ from the hollow cone axis is given by
\begin{eqnarray}
&P_c&={n_e \sigma_T c}U_c \nonumber\\
&&\times \frac{\gamma^2 \beta}{2} \left[\beta\left(3-\frac{K}{J}\right)+2(1+\beta^2)\left(\frac{H}{J}\right)\mu+\beta\left(3\frac{K}{J}-1\right)\mu^2\right]
\label{cone}
\end{eqnarray}
where $J=(\mu_{\rm in}-\mu_{\rm out})$, $H=(\mu^2_{\rm in}-\mu^2_{\rm out})/2$, $K=(\mu^3_{\rm in}-\mu^3_{\rm out})/3$,  $\mu_{\rm in}=\cos\theta_{\rm in}$, $\mu_{\rm out}=\cos\theta_{\rm out}$, $\sigma_T$ is the Thomson cross-section, $n_e$ is the number of electrons, and 
$U_c=(2\pi/c)J \int I_\nu d\nu$ is the radiation energy density of the hollow cone at its apex. Under the head-on approximation, appropriate for relativistic pairs, the bolometric lightcurve is set by the emission of the pairs travelling directly towards the observer. If the pair distribution is isotropic, the lightcurve is given by Eq.~\ref{cone} with $\theta$ representing the angle between the line-of-sight and the cone axis. This angle, which varies from $\pi/2+i$ (superior conjunction) to $\pi/2-i$ (inferior conjunction) with $i$ the system inclination, is calculated using the formulae in \citet{2010A&A...516A..18D}. For comparison, the total power radiated by scattering photons from the star, taken as a point source, is
\begin{equation}
P_\star=n_e \sigma_T c U_\star (1-\beta\mu) \left[(1-\beta\mu)\gamma^2-1\right]
\label{star}
\end{equation}
where $U_\star=(1/c)  \sigma_{SB} T^4_\star (R_\star/d_\star)^2$ with $T_\star$ the star temperature, $R_\star$ its radius, $d_\star$ its distance to the electron. This can be rederived from Eq.~\ref{cone} in the limit $\mu_{\rm in}=1$, $\mu_{\rm out}= 1-(R_\star/d_\star)^2/2\approx 1$, with $\mu\rightarrow -\mu$ because the star is oriented directly opposite to the cone.

\begin{figure}
\centering
\resizebox{\hsize}{!}{\includegraphics{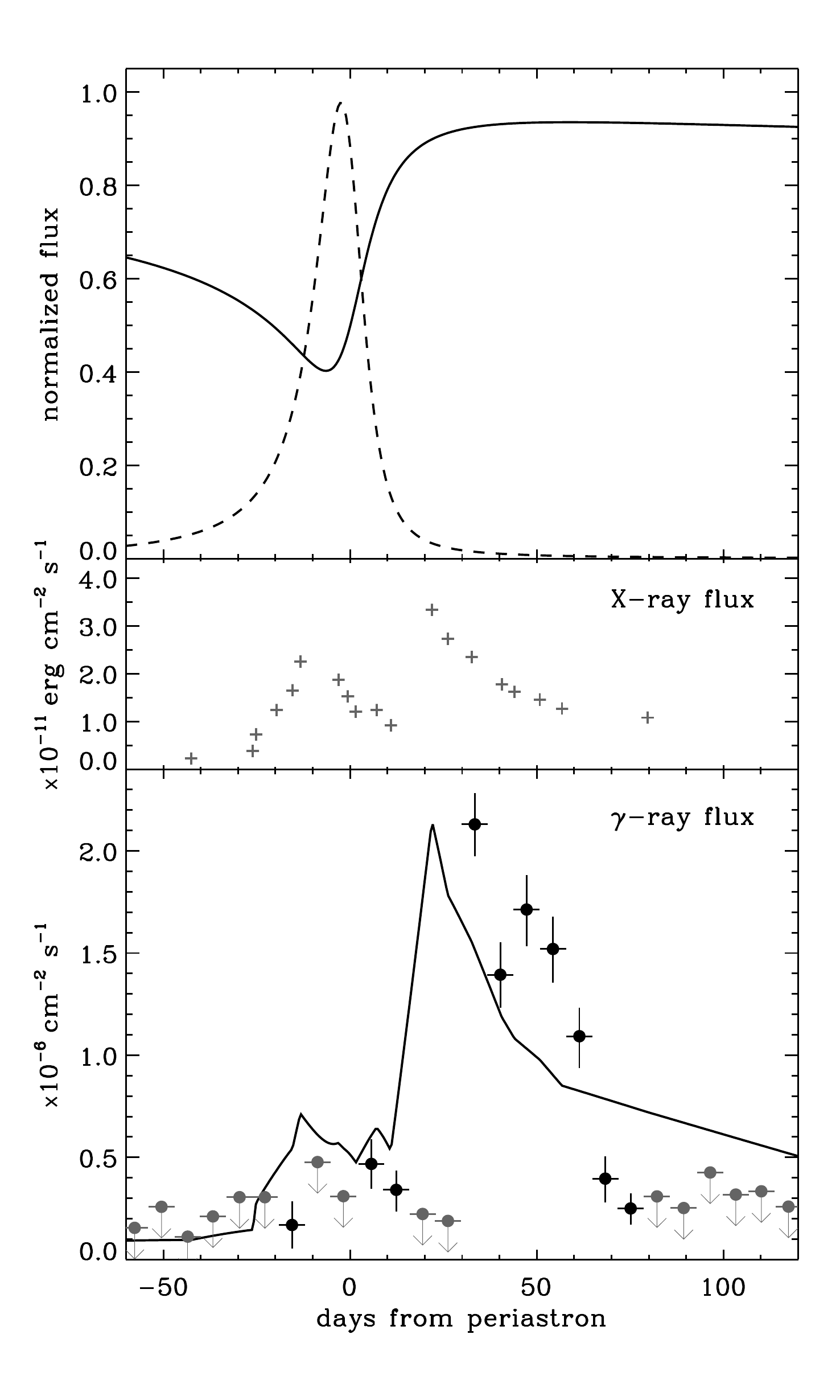}} 
\caption{Top: Lightcurve for inverse Compton scattering on photons from the star (dashed line) and from the hollow shock cone (solid line). Middle: X-ray (1-10\,keV) lightcurve from \citet{2011ApJ...736L..11A}. Bottom: the cone lightcurve is multiplied by the X-ray lightcurve and compared to the  {\em Fermi}/LAT lightcurve. See text for details.}
\end{figure}

\subsection{Results\label{results}}
The lightcurve obtained for scattering off photons from the hollow cone is shown in Fig.~3 (top panel), with $\theta_{\rm in}\approx 50$\degr, $\theta_{\rm out}\approx65$\degr. The lightcurve for scattering off stellar photons is shown for comparison (dashed line). Both are normalized by the same factor $2 n_{e}\sigma_{T}c \gamma^{2} U$ with $U$ constant and taken equal to $U_\star$ at periastron passage (at the pulsar location). The ``cone'' lightcurve peaks at inferior conjunction, as expected, but decays very slowly because the orientation of the cone changes slowly for the observer along this part of the orbit, by a few degrees. The observations show a much faster decline (data in the bottom panel of Fig.~3). Changing the opening angles of the cones has a mild influence on the normalisation or the fractional variability. However, the main characteristics (peak at $\tau+60$\,d and slow decay) remain unchanged.

Other factors must come into play to obtain a better match with the GeV lightcurve. The interaction occurs far from the star, hence the stellar wind has already accelerated to its terminal velocity so $\eta$ is unlikely to change with orbital phase (in as much as we ignore the effects of the Be disc, for the reasons given in \S1). Alternatively, the orientation of the cone could deviate from being radially away from the star. The cone has a large opening angle so parts of it are affected by the Be disc \citep{2012ApJ...750...70T}. However, this effect is unlikely to persist beyond $\tau+60$\,d and, hence, to significantly impact the lightcurve. Finally, the intrinsic emission from the cone could have a strong orbital dependence. The lightcurves shown in the top panel of Fig.~3 assume that $U_{c}$ stays constant along the orbit. Actually, $U_c$ is expected to vary with phase because the changing location and size of the shock region have an impact on the magnetisation at the shock, on the flow timescales, on radiative and adiabatic losses \citep{1996A&AS..120C.221T,Kirk:1999hr,Dubus:2006lc,2007MNRAS.380..320K,2009ApJ...698..911U,2009ApJ...702..100T}. Modelling these effects is beyond the scope of this work. To highlight the possible impact on the lightcurve, we have assumed in the lower panel of Fig.~2 that $U_c$ varies together with the observed X-ray lightcuve (taken from \citealt{2011ApJ...736L..11A}). The impact of the X-ray variations is evident: the geometrical dependence of the inverse Compton cross-section quenches the first X-ray peak and enhances the second X-ray peak in the upscattered lightcurve. Note that both GeV brightening and flare are explained by the same process in this model.

Although the agreement with the observations is improved, the model does not explain the delay between the (second) X-ray peak and the GeV flare. Here, more subtle effects could play a role such as a changing spectral shape for the seed photons, or taking into account a more complex geometry for the location of the high-energy pairs and the seed photons. Doppler boosting may also influence the lightcurve. The shocked pulsar wind accelerates to relativistic velocities away from the apex \citep{2008MNRAS.387...63B} so radiation from particles embedded in the flow will be Doppler boosted in the direction of bulk motion. The effect will be to enhance the emission from the part of the cone moving directly towards the observer, which may slightly modify the lightcurve as our line-of-sight crosses the cone differently with orbital phase. An accurate calculation requires knowledge of the emission and velocity structure inside the cone. However, we note that if the pairs are located at the apex and the X-ray emission is in the wings, where the flow velocity is higher, the seed photon density seen by the pairs will be deboosted, making inverse Compton emission less efficient.

\section{Discussion}
\subsection{Radiative efficiency\label{efficiency}}
The peak gamma-ray luminosity is nearly equal to the pulsar spindown power. The radiative process responsible for the GeV flare must be able to cool the electrons efficiently, before they move away from the gamma-ray emission zone, setting a strict limit on the seed photon energy density  \citep{2012ApJ...752L..17K}. 

We start by considering particles in the unshocked pulsar wind.  Cold particles in the pulsar wind rest frame naturally leads to emission in a narrow band. The pairs travel radially away from the pulsar with $\gamma$ interpreted as the bulk Lorentz factor of the wind $\Gamma$. We see only emission from those pairs that travel directly towards us because of relativistic beaming. They radiate before reaching the termination shock, on a timescale $\tau_{\rm esc}\sim d/v$ with $d$ the orbital separation and $v\approx c$ the speed of the wind. Assuming that the pairs see an isotropic radiation field $U$, their inverse Compton cooling timescale is
\begin{equation}
\tau_{\rm ic}=\frac{3}{4}\frac{m_e c}{\gamma \sigma_T}\frac{1}{U}\approx 6.2\times10^{4}\, \left(\frac{500}{\gamma} \right) \left(\frac{1\, {\rm erg\,cm}^{-3}}{U}\right)\ {\rm s}
\label{tauic}
\end{equation}
assuming the Thomson regime (the timescale using the exact cross-section will be longer by a factor $\sim$ a few at the transition to the Klein-Nishina regime). The lower limit on $U$ is 
\begin{equation}
U\geq 30 \left(\frac{500}{\gamma} \right) \left(\frac{4\,\rm AU}{d}\right)\left(\frac{v}{c}\right)\rm\ erg\,cm^{-3}
\label{uc}
\end{equation}
The pairs do not necessarily see an isotropic field. If they scatter stellar photons then the relevant energy loss rate is given by Eq.~\ref{star}. Since $\theta=60\degr$ at $\tau+60$\,d, the timescale for anisotropic loss rate on stellar photons is $\approx 5$ times longer than given by the isotropic case (Eq.~\ref{tauic}). Stellar photons clearly cannot cool pairs in the unshocked wind since  the required field $U$ is much higher than $U_{\star}\approx 0.1\rm\, erg\,cm^{-3}$ (at the pulsar position and $\tau+60$\,d).  However, if the pairs scatter radiation from the cone, the loss rate given by Eq.~\ref{cone} is 1.4 times smaller than the isotropic case (Eq.~\ref{tauic}), so $U_c\geq 20\rm\ erg\ cm^{-3}$ is a reasonable estimate. The isotropic estimate also holds for the case of particles in the shocked winds because their velocity is isotropized at the shock. Their escape timescale is still roughly estimated as  $\tau_{\rm esc}\sim d/v$, with $v$ the wind speed \citep{Stevens:1992on}. 

The high radiation field density required by efficient cooling is difficult to reconcile with the observed luminosity. The X-ray luminosity is related to $U_c$ in our geometry by 
\begin{equation}
L_X=4\pi U_c c \frac{\int\epsilon_X r^{2}dr}{\int\epsilon_X dr}= 4\pi U_c c \left<r^{2}\right>
\label{lux}
\end{equation}
where $\epsilon_X$ is the X-ray emissivity of the cone, which we assume to depend only on the radius $r$ from apex and to be isotropic. $\left<r^2\right>$ is typically the inner radius of the X-ray emission zone $r_{in}^2$ when $\epsilon_X$ decreases more steeply than $r^{-2}$. The characteristic size is $3\times 10^{10}\ (\gamma/500)^{1/2}(d/4\,{\rm AU})^{1/2} (c/v)^{1/2}\rm\,cm$ since the observed X-ray flux translates to a luminosity $L_X\approx 10^{34}\rm\,erg\,s^{-1}$. The estimated size is very small compared to the natural scale $d$. 

\citet{2012ApJ...752L..17K} face the same difficulty. They considered electrons in the unshocked pulsar wind scattering optical or IR photons, hence they have $\gamma\approx 10^{4}$. The required density $U\geq 1\rm\, erg\,cm^{-3}$ (Eq.~\ref{uc}) is 10 times greater than the stellar radiation density at $\tau+60$\,d. If this is Be disc material heated and disrupted by the pulsar passage, they find its luminosity should be as large as the stellar luminosity (Eq.~\ref{lux} with $r\approx d$). The limit is even greater if the seed photon field is seen behind the gamma-ray emitting electrons by the observer, leading to inefficient Compton scattering, so part of this material must be on our line-of-sight without preventing the detection of radio pulses.

\subsection{Location of the high-energy particles}

Efficient inverse Compton cooling requires a high radiation energy density $U$, posing a major caveat on such models --- including the present one. We propose that this favours locating the gamma-ray emitting particles in the post-shock region, where higher $U$ may be reached, rather than in the pulsar wind. The longer advection timescale means that the particles have more time to radiate, especially if they are mixed with the shocked stellar wind which has $v/c\approx 0.003$. Particles at the apex, near the ``stagnation point'', may take longer to escape than we assume. Our simplified geometry and other assumptions (scattering of X-rays, $U$ independent of exact electron location, isotropic emissivity) also entail significant uncertainties. All of this alleviates the discrepancy in size, although it is unclear whether they can be fully reconciled. Numerical simulations including radiation are required to quantify this.

The GeV-emitting particles could be accelerated at the termination shock of the stellar wind as proposed by \citet{2011MNRAS.418L..49B}. However, this should produce a power-law distribution of particles instead of the narrow distribution of HE particles suggested by observations. Instead, we propose to associate the HE particles with the Maxwellian distribution at energy $\Gamma m_e c^2$ that models of magnetized pair outflows typically produce at the pulsar wind termination shock. In particular, \citet{2011ApJ...741...39S} found in their simulations that shock-driven reconnection of a striped pulsar wind produces thermal distributions at high latitudes, where there is a net magnetic field averaged over a stripe wavelength, and broad non-thermal distributions in the equatorial plane of the pulsar, where the net field averages to zero and particles undergo Fermi acceleration. These broad distributions are obtained for values of $4\pi \kappa R_{LC}/R_{TS}\ga 10$, which is verified here since the termination shock $R_{TS}$ is at  $10^4$-$10^5$ times the light cylinder radius $R_{LC}$. Hence, we speculate that the GeV emission and the X-ray/TeV emission arise from particles situated at different pulsar latitudes in the post-shock region. In any case, assuming as we did that the pairs upscatter  X-ray photons from the shocked flow has for consequence a much lower bulk velocity $\Gamma$ than usually invoked at the termination shock of pulsars like the Crab. There may not be enough room in PSR B1259-63 to accelerate the wind to higher Lorentz factors before it meets its fate at the termination shock.

\subsection{An MeV flare}
A consequence of the model that we propose is that the GeV flare should be preceded by a MeV flare. The high energy pairs responsible for the GeV flare have a Lorentz factor $\gamma\approx 500$ (see \S\ref{model}). These pairs can also upscatter the $\epsilon_\star\approx 10$\,eV stellar photons, resulting in emission peaked at $\approx$ 2 MeV. The lightcurve of this component behaves like the dashed curve in Fig.~3 (top panel), peaking a few days before periastron. The luminosity depends on $n_e$. The density of particles is $n_e\approx 7\times 10^{43}\ U_c^{-1}\rm\,cm^{-3}$ to reproduce the peak GeV flare luminosity $L_{\rm GeV}\approx \dot{E}$. Assuming $n_e$ stays roughly at the same value during periastron passage, using Eq.~\ref{star} gives a peak MeV luminosity  $L_{\rm MeV} \la 4\times 10^{34} \rm\,erg\,s^{-1}$ for  $U_c\geq 30\rm\,erg\,cm^{-3}$. Since this is comparable to $L_X$, a hard component may therefore emerge in hard X-rays beyond 100 keV in the days preceding periastron passage, something that can be tested in the future by the soft gamma-ray detector onboard {\em ASTRO-H}.

\section{Conclusion}
We have explored the possibility to generate a HE gamma-ray flare by inverse Compton scattering X-ray photons emitted by the shocked pulsar wind instead of optical photons from the star. The main advantage of this model is that the associated lightcurve naturally peaks after periastron passage, when the cone of shocked material passes through the line-of-sight, while scattering on stellar or Be disc photons produces a peak before periastron passage. The evolution of the intrinsic emission of the cone with orbital phase needs to be taken into account to reproduce the HE gamma-ray lightcurve, notably the fast decline after $\tau+60$\,d. The high gamma-ray luminosity suggests a high radiative efficiency. As with all models invoking inverse Compton emission, we find that a significant drawback is the very high energy density of seed photons required to have a high radiative efficiency. We speculate that this may be easier to attain in the shock region. The same model could be at work in other gamma-ray binaries, notably LSI +61\degr303 where the GeV emission peaks after periastron passage \citep{2012ApJ...749...54H}.

Finally, we remark that all the proposed models for the GeV flare have tied it to the orbital motion of the pulsar. Hence, all predict a similar GeV flare should occur at the next periastron passage. Its absence, or the detection of a GeV flare at some other orbital phase, could indicate that it was a random occurrence. In this case, the GeV flare of \psrb\ would be more akin to the GeV flares observed in the Crab pulsar wind nebula \citep{2012ApJ...749...26B}, which have been interpreted as random reconnection events downstream the termination shock \citep{2011ApJ...737L..40U,2013ApJ...770..147C}. This is a tempting analogy, although we note that the duration and the radiative efficiency of the flare in PSR~B1259-63 are quite different from the day- to week-long Crab flares, where at most $1\%$ of the spin-down luminosity is radiated away. On the other hand, there is no obvious reason to expect strictly similar particle acceleration and emission from the AU-scale nebula in PSR~B1259-63 and the 0.1~pc nebula in the Crab.

\begin{acknowledgements} We thank Gilles Henri for pointing out that the scattered power (Eq. 2) can be cast in analytical form. GD  acknowledges support from the European Union via contract ERC-StG-200911 and from CNES.
\end{acknowledgements}
\bibliographystyle{aa}
\bibliography{../BIBLIO}

\end{document}